\documentclass[preprint,preprintnumbers,amsmath,amssymb,prl,superscriptaddress]{revtex4-1}

\usepackage{physics}
\usepackage{graphicx}
\usepackage{dcolumn}
\usepackage{bm}
\usepackage{epstopdf}
\usepackage[colorinlistoftodos]{todonotes}
\usepackage{float}
\usepackage[T1]{fontenc}
\usepackage[utf8]{inputenc}
\usepackage{parskip}
\usepackage{physics}
\usepackage{mathtools}

\newcommand*\chem[1]{\ensuremath{\mathrm{#1}}}

\begin{document}

\title{Spectroscopy of bulk and few-layer superconducting NbSe$_2$ with van der Waals tunnel junctions}

\author{T. Dvir}
\affiliation{The Racah Institute of Physics, the Hebrew University of Jerusalem, Israel}

\author{F. Massee}
\affiliation{Laboratoire de Physique des Solides (CNRS UMR 8502), Bâtiment 510, Université Paris-Sud/Université Paris-Saclay, 91405 Orsay, France}

\author{L. Attias}
\affiliation{The Racah Institute of Physics, the Hebrew University of Jerusalem, Israel}

\author{M. Khodas}
\affiliation{The Racah Institute of Physics, the Hebrew University of Jerusalem, Israel}

\author{M. Aprili}
\affiliation{Laboratoire de Physique des Solides (CNRS UMR 8502), Bâtiment 510, Université Paris-Sud/Université Paris-Saclay, 91405 Orsay, France}

\author{C. H. L. Quay}
\affiliation{Laboratoire de Physique des Solides (CNRS UMR 8502), Bâtiment 510, Université Paris-Sud/Université Paris-Saclay, 91405 Orsay, France}

\author{H. Steinberg}
\affiliation{The Racah Institute of Physics, the Hebrew University of Jerusalem, Israel}

\date{\today}

\maketitle

\textbf{Tunnel junctions, a well established platform for high resolution spectroscopy of superconductors, require defect-free insulating barriers with clean engagement to metals on both sides. Extending the range of materials accessible to tunnel junction fabrication, beyond the limited selection which allows high quality oxide formation, requires the development of alternative fabrication techniques. Here we show that van-der-Waals (vdW) tunnel barriers, fabricated by stacking layered semiconductors on top of the transition metal dichalcogenide (TMD) superconductor NbSe$_2$, sustain a stable, low noise tunneling current, and exhibit strong suppression of sub-gap tunneling. We utilize the technique to measure the spectra of bulk (20 nm) and ultrathin (3- and 4-layer) devices at 70 mK. The spectra exhibit two distinct energy gaps, the larger of which decreases monotonously with thickness and $T_C$, in agreement with BCS theory. The spectra are analyzed using a two-band model modified to account for depairing. We show that in the bulk, the smaller gap exhibits strong depairing in an in-plane magnetic field, consistent with a high Fermi velocity. In the few-layer devices, depairing of the large gap is negligible, consistent with out-of-plane spin-locking due to Ising spin-orbit coupling. Our results demonstrate the utility of vdW tunnel junctions in mapping the intricate spectral evolution of TMD superconductors over a range of magnetic fields.
}

%


Superconductors of the transition metal dichalcogenide (TMD) family have seen a revival of interest subsequent to developments in device fabrication by mechanical exfoliation~\cite{Xi_2016, Tsen2015a, Ugeda2015a, Staley2009a, Lu_MoS2_2015, Saito_MoS2_2015, Lu_2017}. The isolation of ultrathin~\chem{NbSe_2}~\cite{Tsen2015a, Xi_2016} has yielded indications of a Berezinskii-Kosterlitz-Thouless transition, which occurs in 2D superconductors. \chem{NbSe_2} \cite {Xi_2016}, gated \chem{MoS_2} \cite{Saito_MoS2_2015,Lu_MoS2_2015} and gated \chem{WS_2}~\cite{Lu_2017} devices also remain superconducting in in-plane magnetic fields well beyond the Pauli limit $H_p = \Delta/(\sqrt{g}\mu_B)$~\cite{Clogston_1962,Chandrasekhar_1962}. 
(Here, $\Delta$ is the superconducting energy gap, $g$ is the Landé g-factor and $\mu_B$ the Bohr magneton.) This is likely due to Ising spin-orbit coupling (ISOC): The broken inversion symmetry of the monolayer TMD in the plane is expected to lead to the formation of Cooper pairs whose constituent spins are locked in the out-of-plane direction, in a singlet configuration.
Interestingly, zero-resistance states have been observed in parallel magnetic fields exceeding the Pauli limit even in few-layer devices, where inversion symmetry is recovered~\cite{Xi_2016}. This suggests that the inter-layer coupling is not strong enough to overcome the out-of-plane spin-locking due to ISOC~\cite{Xi_2016}, perhaps in part due to the presence of spin-layer locking~\cite{Jones2014}. 


These previous studies [1-7] used in-plane electronic transport at high magnetic field and temperatures close to $T_c(H=0)$ to determine the upper critical field $H_{c2\parallel}$, which depends on the magnitude of the spin-orbit field $H_{SO}$. Tunneling spectroscopy can provide complementary information. For example, tunneling and other probes (see Ref.~\cite{Noat2015} and references therein) suggest that (bulk) \chem{NbSe_2} is a 2-band superconductor. The role of the 2$^{nd}$ band, and its response to magnetic field, can be addressed by tunnel spectroscopy but not transport measurements. Tunneling can also probe the effect of magnetic field at the full temperature range and in fields ranging from zero to $H_{c2\parallel}$. 


To carry out tunneling measurements on devices of variable thickness, at variable magnetic field conditions, it is necessary to develop a device-based architecture suitable for integration with TMDs. Oxide-based tunnel barriers, such as those used since the days of Giaever~\cite{Giaever1960a}, have now reached technological maturity; however, there is a limited number of oxides which form high-quality insulating, non-magnetic barriers, and they do not grow well on all surfaces. It is therefore of interest to explore alternatives based on van der Waals (vdW) materials~\cite{Geim2013}, ultrathin layers of which can be precisely positioned on many surfaces~\cite{Dean2010}. Indeed, such barriers have proven effective when integrated with graphene~\cite{Amet2012,Britnell2012,Chandni2016} and appear to be promising candidates for integration with TMD superconductors \cite{island2016thickness}. 



\begin{figure*}
	\centering
	\includegraphics[width=10cm]{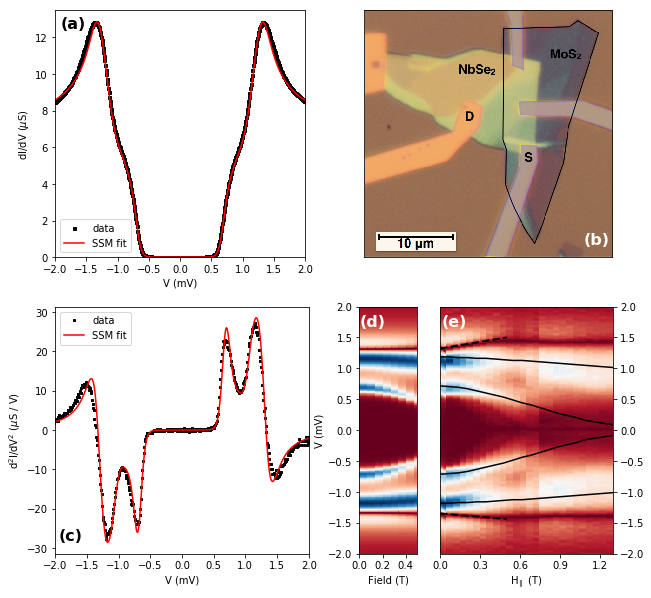}
	\caption[caption]{\textbf{Differential conductance of a bulk NbSe$_2$ tunnel device. a,}  $dI/dV$ vs. $V$ as measured on the device shown in panel b (black) and a fit to the SSM model (red, see details in the text, all fit parameters are given in Supplementary Section 3).  \textbf{b,} Optical image of the tunnel junction device. The yellow-green flake is a 50-20 nm thick \chem{NbSe_2}  (20 nm at the source electrode) and the purple-blue flake is a 4-5 layer \chem{MoS_2} (marked by black outline). Au electrodes are deposited on the left to serve as ohmic contacts (yellow) and on the right to serve as tunnel electrodes (purple).  \textbf{c, } $d^2I/dV^2$  of the data in (a), and the fit to the SSM model. 
\textbf{d,} simulated $|d^2I/dV^2|$ as a function of bias voltage and in-plane magnetic field, in the modified SSM model including a field dependent depairing parameter $\Gamma^{AG}_1 = 0.05~ \textrm{meV/T}^2 $, $\Gamma^{AG}_2 = 0.64~ \textrm{meV/T}^2 $  \textbf{e, } measured $|d^2I/dV^2|$ as a function of in-plane field and voltage bias. The black solid lines follow the positions of the peaks as calculated from the model shown in panel d. The black dashed lines track position of the quasiparticle peak in field, given by  $d^2I/dV^2=0$. The average slope of these lines is 0.27 meV/T. }
	\label{figure1}
\end{figure*}

In this work, we fabricate Normal-Insulator-Superconductor (NIS) tunnel junctions with either \chem{MoS_2} or \chem{WSe_2} --- both vdW materials --- as the insulating barrier. The barrier material is placed on top of \chem{2H-NbSe_2} (hereafter \chem{NbSe_2}), a vdW superconductor with $T_c\approx 7.2$ K, using the “dry transfer” fabrication technique~\cite{Geim2013,Dean2010}. 
Crucially, such heterostructures can comprise \chem{NbSe_2} flakes of varying thicknesses, from `bulk' ($\gtrsim$ 6 layers) to few-layer, often within a single device. Fig. \ref{figure1}b shows a typical junction consisting of a 20nm-thick \chem{NbSe_2} flake partially covered by a 4-5 layer thick \chem{MoS_2} barrier (cf. Supplementary Section 2). The junction has an area $A =$ 1.6 $\mu m^2$, and  we evaluate its transparency to be  $\mathcal{T}\sim10^{-8}$ (Supplementary Section 5).  Panel (a) shows the differential conductance $G = dI/dV$ as a function of $V$ obtained with the device, at $T = 70 mK$. This spectrum has two striking features: first, the very low sub-gap conductance ($G_0 R_N \approx 1/500$), which indicates all bands crossing the Fermi energy are fully gapped. 
Second, the intricate structure of the quasiparticle peak differs from a standard BCS density-of-states (DOS) by having a relatively low  peak and a shoulder at lower energies. The latter feature can be clearly seen in the second derivative (panel c) where the slope separates into a double peak feature, similar to STS scans of bulk \chem{NbSe_2} \cite{Guillamon2008,Noat2015}. Based on this similarity, this flake can be considered bulk in terms of the zero field superconducting properties, and is hence referred to as the `bulk' sample.
 

%
%

Density functional theory calculations~\cite{Johannes2006}, and ARPES data~\cite{Yokoya2001a} show that the dispersion of \chem{NbSe_2} consists of 5 independent bands which cross the Fermi energy. Of these, four are Nb-derived bands with roughly cylindrical Fermi surfaces, centered at the $\Gamma$ and $K$ points. The fifth is derived from the Se $p_z$ orbitals, which give rise to a small ellipsoid pocket around the $\Gamma$ point. Ref~\cite{Noat2015} uses a two-band model to fit \chem{NbSe_2} tunneling data, which can be justified noting that the Se and Nb-derived bands differ in the density of states and value of the electron-phonon coupling parameter~\cite{Kiss2007}. We follow Ref~\cite{Noat2015} in fitting our data using the same two-band model, which was developed in various forms by Suhl~\cite{Suhl1959}, Schopohi~\cite{Schopohi1977a} and McMillan~\cite{McMillan1968} (below `SSM'). 

The model entails a self-consistent solution to the coupled equations for the energy dependent order parameters $\Delta_i (E)$ in the two bands $i$:

\begin{eqnarray}
\Delta_i (E) = \frac{\Delta_i^0 + \Gamma_{ij} \Delta_j(E)/\sqrt{\Delta_j^2(E)-E^2}}{1 + \Gamma_{ij} /\sqrt{\Delta_j^2(E)-E^2}+\Gamma_i^{AG}/\sqrt{\Delta_i^2(E)-E^2}}
\label{KZQ}
\end{eqnarray} 
$\Delta_i^0$ describes the intrinsic gap within each band $i$, that is generated by the electron-phonon coupling and by the scattering rates of quasiparticles between the bands $\Gamma_{ij}$. The extension of the two-band model to include Abrikosov-Gor'kov depairing~\cite{AbrikosovA.A.1961ContributionImpurities,Maki1964TheCurrents, Levine1967DensityTunneling, Millstein1967TunnelingField} 
 --- via the terms with $\Gamma_i^{AG}$ ---  was done by Kaiser and Zuckermann~\cite{Keiser1970}. Here, depairing is due to magnetic field; thus, $\Gamma_i^{AG}$ are set to 0 when no magnetic field is applied. The DOS of each band is then given by
\begin{eqnarray}
N_S^i(E)  = N_i(E_F) \frac{1}{2 \pi} \int d\theta \Re{\frac{|E|}{\sqrt{ (1 + \alpha \cos \theta) \Delta_i^2(E)-E^2}}} ,
\end{eqnarray}
where $N_i(E_F)$ is the DOS at the Fermi energy in the normal state in band $i$. The parameter $\alpha = 0.1$, incorporates band-anisotropy. This anisotropy also affects the effective fit temperature ($T = 0.44\ \mathrm{K}$), which is higher than the sample temperature.
Our fit indicates the presence of two independent order parameters. The larger,  $\Delta_1^0 = 1.26\pm0.01\  \mathrm{meV}$ can be determined with high fidelity. The smaller order parameter $\Delta_2^0$, can not be determined unambiguously, and could have any value between 0 and 0.3 meV. The suppressed intrinsic superconductivity in the second band is consistent with a band with small density of states and weak electron-phonon coupling. The Se-band, having these properties~\cite{Kiss2007}, is thus a candidate for the 2nd band.

We next investigate the evolution of the tunneling spectra vs. in-plane magnetic field $H_\parallel$ (Figure~\ref{figure1}e). 
The bulk sample appearing in Figure~\ref{figure1} is thick enough to accommodate Meissner currents leading to orbital depairing. We track the peaks in $d^2I/dV^2$, which are associated with the two gaps. The high energy peak in $d^2I/dV^2$ depends weakly on the field, whereas the low energy peak evolves nonlinearly towards lower energies. This trend persists up to $H = 0.5T$ which we interpret as $H_{C1}$. 
Although \chem{NbSe_2} is a clean-limit superconductor (coherence length $\xi < l$ mean free path), the evolution of the low energy gap in $H_\parallel$ does not agree with the relevant, Doppler shift model~\cite{Fulde1969book}. Instead, the magnetic field evolution of its $d^2I/dV^2$ features towards lower energies is well-reproduced by the diffusive Kaiser-Zuckermann (KZ) model (Figure~\ref{figure1}d), assuming a depairing parameter $\Gamma^{AG}_2$ quadratic in $H_\parallel$. For a thin sample of thickness $d << \lambda$,  $\Gamma_i^{AG}=D_ie^2H_{\parallel}^2d^2/6\hbar c^2$~\cite{tinkham1996introduction}, with $D_i$ the diffusion coefficient, $d \approx 20$ nm and the penetration length $\lambda\approx$ 230 nm~\cite{{Garoche1976a}}. We find $\Gamma^{AG}_2 =$ 640~$\mu$eV at 1~T, corresponding to $D_2 =$ 40~cm$^2$/s. The model therefore yields a very large value for the diffusion coefficient $D_2=v_Fl/3$, indicating a large Fermi velocity. This lends further support to the identification of this feature with the Se-derived band, where $v_F$ is 4-5 times larger than in the Nb-derived bands~\cite{Kiss2007}. In contrast, the high energy quasiparticle peak moves only slightly and linearly to higher energies at low $H_\parallel$ (dashed line in Figure \ref{figure1}e). From this, we estimate $v_F \approx 5\times 10^4$~m/s. This indicates that a comprehensive description of the field evolution of the full spectrum would require a model with arbitrary disorder bridging clean and diffusive limits. 

\begin{figure} 
\centering
\includegraphics[width = 0.5\textwidth]{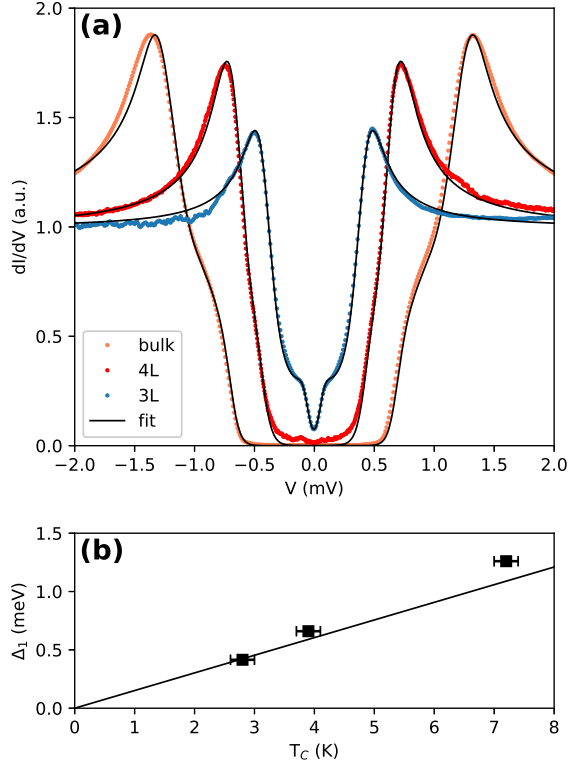}
\caption{\textbf{Thickness dependent tunneling spectrum. 
}   \textbf{a,} Normalized differential conductance of vdW tunnel junctions with \chem{NbSe_2} of varying thickness: 20nm thick (bulk, orange), 4 layers  (red) and 3 layers  (blue). Each curve is fit to the SSM model (fit parameters reported in Supplementary Section 3).\textbf{b,} $\Delta_1^0$ of the devices in (a), extracted from the SSM model, as a function of $T_c$. Solid line shows a comparison to the BCS prediction: $\Delta = 1.76 K_B T_c $.
}
 \label{Thin_dIdV}
\end{figure}

Our methods allow us to carry out a straightforward comparison of the tunneling spectra from devices with different \chem{NbSe_2} thicknesses. Figure \ref{Thin_dIdV}a shows differential conductance curves taken by tunneling into ultra-thin \chem{NbSe_2} flakes (3 layers, 4 layers), in comparison to the $dI/dV$ of the bulk flake discussed above~\footnote{The 4-layer junction includes a small trilayer, but this has no apparent effect on the tunneling spectrum.}. The spectra of the thin devices are in good agreement with the SSM model. 
We extract the values of $\Delta_1^0$ from these fits, and separately evaluate $T_c$ using the temperature dependence of the tunneling conductance (details in Supplementary Section 5). We find that $\Delta_1^0$ increases with $T_c$ (Figure \ref{Thin_dIdV}b), however, deviating slightly from the BCS result $\Delta = 1.76 k_B T_c$. This deviation is plausible due to the multi-band nature of superconductivity in \chem{NbSe_2}.
The $T_c$ values measured here are lower than those reported elsewhere~\cite{Xi_2016,Frindt}, likely due to higher disorder in our sample ~\cite{Finkelshtein1987,Finkelshtein1994,Goldman}. Note that, based on Ref.~\cite{Frindt}, it would seem that the $T_c$ dependence on number of layers seen by all works to date is not due to strain or other substrate effects.


\begin{figure*}
\includegraphics[width =\linewidth]{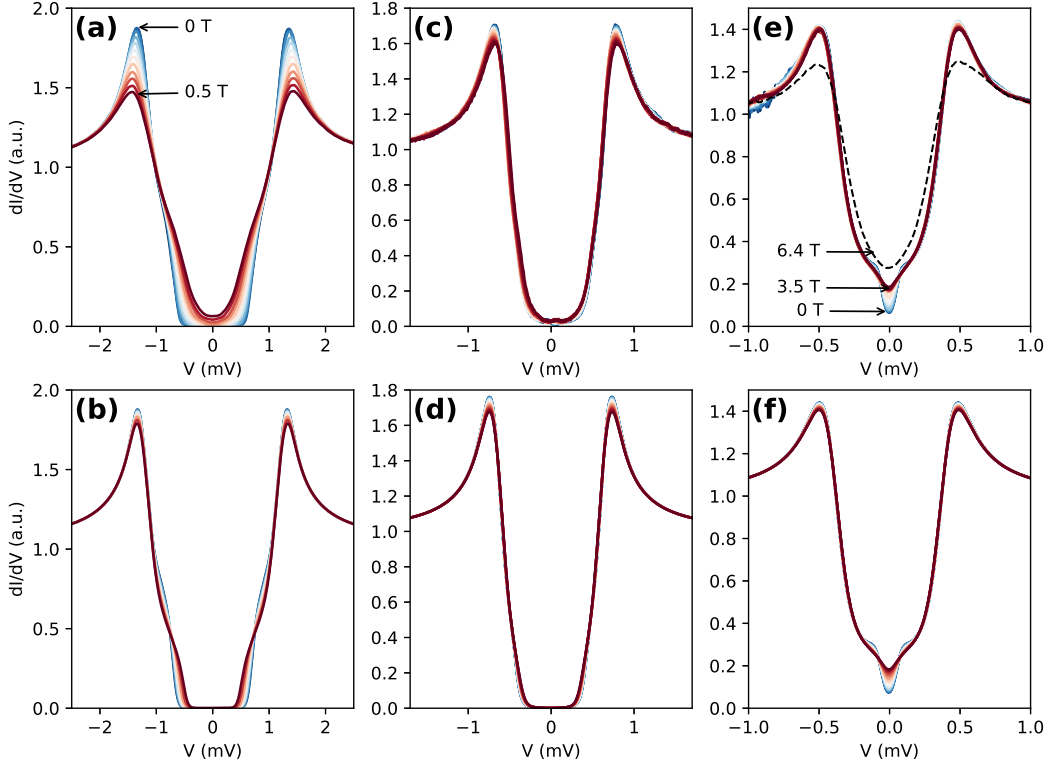}
\caption{\textbf{Response of the tunneling conductance to parallel magnetic fields. 
a,} $dI/dV$ curves at increasing magnetic field parallel to the \chem{NbSe_2} layers  ($H_\parallel$) of the bulk sample, at $0 < H_\parallel < 0.5$T \textbf{b,} Same for the 4-layer device at $0 < H_\parallel < 3.5$T, and \textbf{c,} same for the 3-layer  device.  Dashed line in (c): $dI/dV$ curve taken with the 3-layer device at uncompensated parallel field of 6.4T.  
\textbf{d, e, f} $dI/dV$ calculated from the KZ model for the bulk, 4- layer and 3- Layer respectively. Model parameters are given in supplementary section 3. 
}
\label{F_ThinField}
\end{figure*}

We now turn to the response of the ultrathin devices to in-plane magnetic fields. 
Figure~\ref{F_ThinField} shows the tunneling spectra for the bulk device (panel a) and for the 4-layer and 3-layer devices (panels b and c, respectively).  In the latter, unlike the bulk device, we find that the spectrum changes very little up to 3.5T (which is the maximal field where a compensation coil keeps a zero perpendicular field). In both 3 and 4 layer devices, there is a small reduction in the height of the quasiparticle peak; in the 3 layer device, the low energy spectrum exhibits a more intricate evolution (discussed below). Using the KZ model to quantify the reduction of the peak height, we find that the depairing term $\Gamma_1^{AG} \approx 0.5\mu$ eV at 1T.

Since orbital depairing is quadratic in sample thickness, we expect it to be diminished in the 3 and 4 layer devices, allowing us to probe the spin-dependent interaction.  The interaction of the spin with magnetic field should lead to Zeeman splitting of the spectrum, which we do not see. This could be due to two mechanisms. First, spin-orbit scattering can effectively randomize the spin, giving rise to a depairing parameter given by $\Gamma^{AG}\ = \tau_{SO}e^2\hbar H_{\parallel}^2/2m^2$, where $\tau_{SO} $ is the spin orbit scattering time~\cite{tinkham1996introduction}. Second, ISOC can align the spins in the out-of-plane direction with an effective field, $H_{SO}$, and the depairing term $\Gamma^{AG} \sim 2 \mu_B H_\parallel^2/H_{SO}$ \cite{Lu_MoS2_2015,Xi_2016}. The first scenario can be ruled out, since it yields $\tau_{SO}<$ 50~fs, shorter than the scattering time~\cite{Lu_MoS2_2015}. The ISOC case is more likely, and the depairing energy of $0.5\mu eV$ at 1T (for the 3L device) yields $H_{SO} \approx 200$~T. Using $H_p = 4.9T$ (extracted by setting $\Delta_1 = 0.4$meV) we can estimate $H_{c2}=\sqrt{H_{SO}H_p}\approx$30T, consistent with transport experiments \cite{Xi_2016}. We note that since the orbital term is not entirely suppressed, and we cannot estimate its contribution to the depairing, the estimate for $H_{SO}$ is a lower bound. Further details concerning the possible interpretation of the depairing term are given in supplementary section 3.

The stability of the larger gap can even be demonstrated above the Pauli limit, $H_p = 4.9T$ for the 3 layer device. In Figure~\ref{F_ThinField}c we present the density of states of this sample above the Pauli limit by applying an in-plane field of 6.4T. At this field, our measurement system did not allow us to compensate for angle misalignment leading to small component of perpendicular field ($\approx$0.2T) and possible vortex penetration. Nevertheless, the size of the gap remains unchanged. 
This lends further support to ISOC as the mechanism protecting \chem{NbSe_2} superconductivity at high parallel fields.




The sub-gap spectrum of the 3-layer device appears to exhibit a secondary, well-formed small gap of $\Delta_2=50\mu V$, which is suppressed at $H_\parallel = 1.2T$. As we show in Figure~\ref{F_ThinField}(c,f), the KZ model reproduces this data remarkably well. Here, too, the depairing term is quadratic in $H_\parallel$, with $\Gamma^{AG}_2\approx$ 13~$\mu$eV at 1~T. 
This value is to big to be interpreted in terms of orbital depairing. The observed depairing could alternatively be associated with spin-orbit-driven spin-flip scattering, with $\tau_{SO} \approx 1.5$~ps. This depairing shows that unlike the band with larger gap, this band is not protected by ISOC. 
This difference can be explained by associating the smaller order parameter with the Se-derived band. The outer gap, appearing immune to depairing, would then be associated with the Nb-derived $K$-band. We note that this interpretation leaves  open the question of the role of the Nb-derived $\Gamma$-band. Addressing this question will require further spectral studies - in particular, of monolayer \chem{NbSe_2} where the Se-derived band does not cross the Fermi energy.

Our results show that TMD semiconductors transferred on top of \chem{NbSe_2}  form stable tunnel barriers with a hard gap. We show that the SSM model, modified to include diffusive depairing, successfully reproduces the tunneling spectra in both the bulk and the ultrathin limits, at the presence of in-plane magnetic fields. This allows us to probe the effect of the spin and orbital degrees of freedom on the spectra, thereby differentiating between the responses of the different bands to the field. The large gaps, in the 3 and 4 layer devices, are remarkably stable to depairing by the in-plane field, exhibiting very small depairing energies (<~1~$\mu$eV), which place a tight cap on the spin-depairing observed on this band, lending support to ISOC as the mechanism behind this stability. 
We suggest that our technique can be generalized to work with many other material systems, such as organic (super)conductors and other fragile systems which have hitherto not been investigated in tunneling spectroscopy. 

\section*{Methods}
The vdW tunnel junctions were fabricated using the dry transfer technique \cite{Castellanos-Gomez2014}, carried out in a glove-box (nitrogen atmosphere). \chem{NbSe_2} flakes were cleaved using the scotch tape method, exfoliated on commercially available Gelfilm from Gelpack, and subsequently transferred to a \chem{SiO_2} substrate. \chem{MoS_2} and \chem{WSe_2} flakes were similarly exfoliated and thin flakes suitable for the formation of tunnel barriers were selected based on optical transparency. The barrier flake was then transferred and positioned on top of the \chem{NbSe_2} flake at room temperature. Ti/Au contacts and tunnel electrodes were fabricated using standard e-beam techniques. Prior to the evaporation of the ohmic contacts the sample was ion milled for 15 seconds. No such treatment was done with the evaporation of the tunnel electrodes. All transport measurements were done in a $^3$He–$^4$He dilution refrigerator with a base temperature of 70 mK. The AC excitation voltage was modulated at 17 Hz; its amplitude was 15$\mu$V at all temperatures for the bulk device and 10$\mu$V for the few-layer devices. Measurement circuit details are provided in Supplementary Section 1.

\section*{Acknowledgements}
We thank P. Février and J. Gabelli for helpful discussions on tunnel barriers, and T. Cren for the same on \chem{NbSe_2}. This work was funded by a Maimonïdes-Israel grant from the Israeli-French High Council for Scientific \& Technological Research and by an ANR JCJC grant (SPINOES) from the French Agence Nationale de Recherche. H.S. acknowledges support by ERC-2014-STG Grant No. 637298 (TUNNEL) and Marie Curie CIG Grant No. PCIG12-GA-2012-333620. T.D. is grateful to the Azrieli Foundation for an Azrieli Fellowship. F.M. has received funding from the European Union’s Horizon 2020 research and innovation programme under the Marie Skłodowska-Curie Grant No. 659247. L.A. and M.K. are supported by the Israeli Science Foundation through Grant No. 1287/15.

\section*{Author contributions}
T.D. fabricated the devices. C.Q.H.L., T.D. and M.A. performed the measurements. All the authors contributed to data analysis and the writing of the manuscript.

\section*{Competing financial interests}
The authors declare no competing financial interests.

\bibliographystyle{naturemag}
\bibliography{Corrected_bib}

\begin{thebibliography}{10}
\expandafter\ifx\csname url\endcsname\relax
  \def\url#1{\texttt{#1}}\fi
\expandafter\ifx\csname urlprefix\endcsname\relax\def\urlprefix{URL }\fi
\providecommand{\bibinfo}[2]{#2}
\providecommand{\eprint}[2][]{\url{#2}}

\bibitem{Xi_2016}
\bibinfo{author}{Xi, X.} \emph{et~al.}
\newblock \bibinfo{title}{{Ising pairing in superconducting NbSe$_2$ atomic
  layers}}.
\newblock \emph{\bibinfo{journal}{Nature Physics}}
  \textbf{\bibinfo{volume}{12}}, \bibinfo{pages}{139--143}
  (\bibinfo{year}{2016}).

\bibitem{Tsen2015a}
\bibinfo{author}{Tsen, a.~W.} \emph{et~al.}
\newblock \bibinfo{title}{{Nature of the Quantum Metal in a Two-Dimensional
  Crystalline Superconductor}}.
\newblock \emph{\bibinfo{journal}{Nature Physics}} \bibinfo{pages}{1--8}
  (\bibinfo{year}{2015}).

\bibitem{Ugeda2015a}
\bibinfo{author}{Ugeda, M.~M.} \emph{et~al.}
\newblock \bibinfo{title}{{Characterization of collective ground states in
  single-layer NbSe$_2$}}.
\newblock \emph{\bibinfo{journal}{Nature Physics}}
  \textbf{\bibinfo{volume}{12}}, \bibinfo{pages}{92--97}
  (\bibinfo{year}{2015}).

\bibitem{Staley2009a}
\bibinfo{author}{Staley, N.~E.} \emph{et~al.}
\newblock \bibinfo{title}{{Electric field effect on superconductivity in
  atomically thin flakes of NbSe$_2$}}.
\newblock \emph{\bibinfo{journal}{Physical Review B}}
  \textbf{\bibinfo{volume}{80}}, \bibinfo{pages}{1--6} (\bibinfo{year}{2009}).

\bibitem{Lu_MoS2_2015}
\bibinfo{author}{Lu, J.~M.} \emph{et~al.}
\newblock \bibinfo{title}{{Evidence for two-dimensional Ising superconductivity
  in gated MoS$_2$}}.
\newblock \emph{\bibinfo{journal}{Science}} \textbf{\bibinfo{volume}{350}},
  \bibinfo{pages}{1353--1357} (\bibinfo{year}{2015}).

\bibitem{Saito_MoS2_2015}
\bibinfo{author}{Saito, Y.} \emph{et~al.}
\newblock \bibinfo{title}{{Superconductivity protected by
  spin{\textendash}valley locking in ion-gated MoS$_2$}}.
\newblock \emph{\bibinfo{journal}{Nature Physics}}
  \textbf{\bibinfo{volume}{12}}, \bibinfo{pages}{144--149}
  (\bibinfo{year}{2015}).

\bibitem{Lu_2017}
\bibinfo{author}{Lu, J.~M.} \emph{et~al.}
\newblock \bibinfo{title}{{A full superconducting dome of strong Ising
  protection in gated monolayer WS$_2$}}.
\newblock \emph{\bibinfo{journal}{arXiv preprint arXiv:1703.06369}}
  (\bibinfo{year}{2017}).

\bibitem{Clogston_1962}
\bibinfo{author}{Clogston, A.~M.}
\newblock \bibinfo{title}{{Upper Limit for the Critical Field in Hard
  Superconductors}}.
\newblock \emph{\bibinfo{journal}{Physical Review Letters}}
  \textbf{\bibinfo{volume}{9}}, \bibinfo{pages}{266--267}
  (\bibinfo{year}{1962}).

\bibitem{Chandrasekhar_1962}
\bibinfo{author}{Chandrasekhar, B.~S.}
\newblock \bibinfo{title}{{A note on the maximum critical field of high-field
  superconductors}}.
\newblock \emph{\bibinfo{journal}{Applied Physics Letters}}
  \textbf{\bibinfo{volume}{1}}, \bibinfo{pages}{7} (\bibinfo{year}{1962}).

\bibitem{Jones2014}
\bibinfo{author}{Jones, A.~M.} \emph{et~al.}
\newblock \bibinfo{title}{Spin-layer locking effects in optical orientation of
  exciton spin in bilayer $\mathrm{WSe}_2$}.
\newblock \emph{\bibinfo{journal}{Nature Physics}}
  \textbf{\bibinfo{volume}{10}}, \bibinfo{pages}{130--134}
  (\bibinfo{year}{2014}).

\bibitem{Noat2015}
\bibinfo{author}{Noat, Y.} \emph{et~al.}
\newblock \bibinfo{title}{{Quasiparticle spectra of 2H-NbSe$_2$: Two-band
  superconductivity and the role of tunneling selectivity}}.
\newblock \emph{\bibinfo{journal}{Physical Review B}}
  \textbf{\bibinfo{volume}{92}}, \bibinfo{pages}{1--18} (\bibinfo{year}{2015}).

\bibitem{Giaever1960a}
\bibinfo{author}{Giaever, I.}
\newblock \bibinfo{title}{{Energy gap in superconductors measured by electron
  tunneling}}.
\newblock \emph{\bibinfo{journal}{Physical Review Letters}}
  \textbf{\bibinfo{volume}{5}}, \bibinfo{pages}{147--148}
  (\bibinfo{year}{1960}).

\bibitem{Geim2013}
\bibinfo{author}{Geim, A.~K.} \& \bibinfo{author}{Grigorieva, I.~V.}
\newblock \bibinfo{title}{{Van der Waals heterostructures}}.
\newblock \emph{\bibinfo{journal}{Nature}} \textbf{\bibinfo{volume}{499}},
  \bibinfo{pages}{419--425} (\bibinfo{year}{2013}).

\bibitem{Dean2010}
\bibinfo{author}{Dean, C.~R.} \emph{et~al.}
\newblock \bibinfo{title}{{Boron nitride substrates for high-quality graphene
  electronics.}}
\newblock \emph{\bibinfo{journal}{Nature Nanotechnology}}
  \textbf{\bibinfo{volume}{5}}, \bibinfo{pages}{722--726}
  (\bibinfo{year}{2010}).

\bibitem{Amet2012}
\bibinfo{author}{Amet, F.} \emph{et~al.}
\newblock \bibinfo{title}{{Tunneling spectroscopy of graphene-boron-nitride
  heterostructures}}.
\newblock \emph{\bibinfo{journal}{Physical Review B}}
  \textbf{\bibinfo{volume}{85}} (\bibinfo{year}{2012}).

\bibitem{Britnell2012}
\bibinfo{author}{Britnell, L.} \emph{et~al.}
\newblock \bibinfo{title}{{Electron tunneling through ultrathin boron nitride
  crystalline barriers}}.
\newblock \emph{\bibinfo{journal}{Nano Letters}} \textbf{\bibinfo{volume}{12}},
  \bibinfo{pages}{1707--1710} (\bibinfo{year}{2012}).

\bibitem{Chandni2016}
\bibinfo{author}{Chandni, U.}, \bibinfo{author}{Watanabe, K.},
  \bibinfo{author}{Taniguchi, T.} \& \bibinfo{author}{Eisenstein, J.~P.}
\newblock \bibinfo{title}{{Signatures of phonon and defect-assisted tunneling
  in planar metal-hexagonal boron nitride-graphene junctions}}.
\newblock \emph{\bibinfo{journal}{Nano Letters}} \textbf{\bibinfo{volume}{16}},
  \bibinfo{pages}{7982--7987} (\bibinfo{year}{2016}).

\bibitem{island2016thickness}
\bibinfo{author}{Island, J.~O.}, \bibinfo{author}{Steele, G.~A.},
  \bibinfo{author}{van~der Zant, H.~S.} \& \bibinfo{author}{Castellanos-Gomez,
  A.}
\newblock \bibinfo{title}{Thickness dependent interlayer transport in vertical
  mos$_2$ josephson junctions}.
\newblock \emph{\bibinfo{journal}{2D Materials}} \textbf{\bibinfo{volume}{3}},
  \bibinfo{pages}{031002} (\bibinfo{year}{2016}).

\bibitem{Guillamon2008}
\bibinfo{author}{Guillam\'on, I.}, \bibinfo{author}{Suderow, H.},
  \bibinfo{author}{Guinea, F.} \& \bibinfo{author}{Vieira, S.}
\newblock \bibinfo{title}{{Intrinsic atomic-scale modulations of the
  superconducting gap of 2H-NbSe$_2$}}.
\newblock \emph{\bibinfo{journal}{Physical Review B}}
  \textbf{\bibinfo{volume}{77}} (\bibinfo{year}{2008}).

\bibitem{Johannes2006}
\bibinfo{author}{Johannes, M.~D.}, \bibinfo{author}{Mazin, I.~I.} \&
  \bibinfo{author}{Howells, C.~A.}
\newblock \bibinfo{title}{{Fermi-surface nesting and the origin of the
  charge-density wave in NbSe$_2$}}.
\newblock \emph{\bibinfo{journal}{Physical Review B}}
  \textbf{\bibinfo{volume}{73}}, \bibinfo{pages}{205102}
  (\bibinfo{year}{2006}).

\bibitem{Yokoya2001a}
\bibinfo{author}{Yokoya, T.} \emph{et~al.}
\newblock \bibinfo{title}{{Fermi surface sheet-dependent superconductivity in
  2H-NbSe2.}}
\newblock \emph{\bibinfo{journal}{Science}} \textbf{\bibinfo{volume}{294}},
  \bibinfo{pages}{2518--20} (\bibinfo{year}{2001}).

\bibitem{Kiss2007}
\bibinfo{author}{Kiss, T.} \emph{et~al.}
\newblock \bibinfo{title}{Charge-order-maximized momentum-dependent
  superconductivity}.
\newblock \emph{\bibinfo{journal}{Nature Physics}}
  \textbf{\bibinfo{volume}{3}}, \bibinfo{pages}{720--725}
  (\bibinfo{year}{2007}).

\bibitem{Suhl1959}
\bibinfo{author}{Suhl, H.}, \bibinfo{author}{Matthias, B.~T.} \&
  \bibinfo{author}{Walker, L.~R.}
\newblock \bibinfo{title}{{Bardeen-Cooper-Schrieffer Theory of
  superconductivity in the case of overlapping bands}}.
\newblock \emph{\bibinfo{journal}{Physical Review Letters}}
  \textbf{\bibinfo{volume}{3}}, \bibinfo{pages}{552--554}
  (\bibinfo{year}{1959}).

\bibitem{Schopohi1977a}
\bibinfo{author}{Schopohl, N.} \& \bibinfo{author}{Scharnberg, K.}
\newblock \bibinfo{title}{{Tunneling Density of States for the Two-Band Model
  of Superconductivity}}.
\newblock \emph{\bibinfo{journal}{Solid State Communications}}
  \textbf{\bibinfo{volume}{22}}, \bibinfo{pages}{37--1} (\bibinfo{year}{1977}).

\bibitem{McMillan1968}
\bibinfo{author}{McMillan, W.~L.}
\newblock \bibinfo{title}{{Tunneling model of the superconducting proximity
  effect}}.
\newblock \emph{\bibinfo{journal}{Physical Review}}
  \textbf{\bibinfo{volume}{175}}, \bibinfo{pages}{537--542}
  (\bibinfo{year}{1968}).

\bibitem{AbrikosovA.A.1961ContributionImpurities}
\bibinfo{author}{{Abrikosov A.A.}} \& \bibinfo{author}{{Gor'kov L.P.}}
\newblock \bibinfo{title}{{Contribution to the Theory of Superconducting Alloys
  with Paramagnetic Impurities}}.
\newblock \emph{\bibinfo{journal}{Soviet Physics JETP}}
  \textbf{\bibinfo{volume}{12}}, \bibinfo{pages}{1243} (\bibinfo{year}{1961}).

\bibitem{Maki1964TheCurrents}
\bibinfo{author}{Maki, K.}
\newblock \bibinfo{title}{{The Behavior of Superconducting Thin Films in the
  Presence of Magnetic Fields and Currents}}.
\newblock \emph{\bibinfo{journal}{Progress of Theoretical Physics}}
  \textbf{\bibinfo{volume}{31}}, \bibinfo{pages}{731--741}
  (\bibinfo{year}{1964}).

\bibitem{Levine1967DensityTunneling}
\bibinfo{author}{Levine, J.~L.}
\newblock \bibinfo{title}{{Density of States of a Short-Mean-Free-Path
  Superconductor in a Magnetic Field by Electron Tunneling}}.
\newblock \emph{\bibinfo{journal}{Physical Review}}
  \textbf{\bibinfo{volume}{155}}, \bibinfo{pages}{373} (\bibinfo{year}{1967}).

\bibitem{Millstein1967TunnelingField}
\bibinfo{author}{Millstein, J.} \& \bibinfo{author}{Tinkham, M.}
\newblock \bibinfo{title}{{Tunneling into superconducting films in a magnetic
  field}}.
\newblock \emph{\bibinfo{journal}{Physical Review}}
  \textbf{\bibinfo{volume}{158}}, \bibinfo{pages}{325--332}
  (\bibinfo{year}{1967}).

\bibitem{Keiser1970}
\bibinfo{author}{Kaiser, A.~B.} \& \bibinfo{author}{Zuckermann, M.~J.}
\newblock \bibinfo{title}{{McMillan Model of the Superconducting Proximity
  Effect for Dilute Magnetic Alloys}}.
\newblock \emph{\bibinfo{journal}{Physical Review B}}
  \textbf{\bibinfo{volume}{1}}, \bibinfo{pages}{229--235}
  (\bibinfo{year}{1970}).

\bibitem{Fulde1969book}
\bibinfo{author}{Fulde, P.}
\newblock \emph{\bibinfo{title}{Tunneling phenomena in solids}}, chap.
  \bibinfo{chapter}{Gapless Superconducting Tunneling- Theory}
  (\bibinfo{publisher}{Springer}, \bibinfo{year}{1969}).

\bibitem{tinkham1996introduction}
\bibinfo{author}{Tinkham, M.}
\newblock \emph{\bibinfo{title}{Introduction to superconductivity}}
  (\bibinfo{publisher}{Courier Corporation}, \bibinfo{year}{1996}).

\bibitem{Garoche1976a}
\bibinfo{author}{Garoche, P.}, \bibinfo{author}{Veyssi\'e, J.},
  \bibinfo{author}{P, M.} \& \bibinfo{author}{P, M.}
\newblock \bibinfo{title}{{Experimental inverstigation of superconductivity in
  2H-NbSe$_2$ single crystal}}.
\newblock \emph{\bibinfo{journal}{Solid State Communications}}
  \textbf{\bibinfo{volume}{19}}, \bibinfo{pages}{455--460}
  (\bibinfo{year}{1976}).

\bibitem{Note1}
\bibinfo{note}{The 4-layer junction includes a small trilayer, but this has no
  apparent effect on the tunneling spectrum.}

\bibitem{Frindt}
\bibinfo{author}{Frindt, R.~F.}
\newblock \bibinfo{title}{{Superconductivity in Ultrathin NbSe$_2$ Layers}}.
\newblock \emph{\bibinfo{journal}{Physical Review Letters}}
  \textbf{\bibinfo{volume}{5}}, \bibinfo{pages}{299--301}
  (\bibinfo{year}{1972}).

\bibitem{Finkelshtein1987}
\bibinfo{author}{Finkel'shtein, A.~M.}
\newblock \bibinfo{title}{{Superconducting transition temperature in Amorphous
  Films}}.
\newblock \emph{\bibinfo{journal}{JETP}} \textbf{\bibinfo{volume}{45}},
  \bibinfo{pages}{37--40} (\bibinfo{year}{1987}).

\bibitem{Finkelshtein1994}
\bibinfo{author}{Finkel'shtein, A.~M.}
\newblock \bibinfo{title}{{Suppression of superconductivity in homogeneously
  disordered systems}}.
\newblock \emph{\bibinfo{journal}{Physica B}} \textbf{\bibinfo{volume}{197}},
  \bibinfo{pages}{636--648} (\bibinfo{year}{1994}).

\bibitem{Goldman}
\bibinfo{author}{Goldman, A.~M.} \& \bibinfo{author}{Markovi\'c, N.}
\newblock \bibinfo{title}{{Superconductor‐insulator transitions in the
  two‐dimensional limit}}.
\newblock \emph{\bibinfo{journal}{Physics Today}}
  \textbf{\bibinfo{volume}{51}}, \bibinfo{pages}{39--44}
  (\bibinfo{year}{1998}).

\bibitem{Castellanos-Gomez2014}
\bibinfo{author}{Castellanos-Gomez, A.} \emph{et~al.}
\newblock \bibinfo{title}{{Deterministic transfer of two-dimensional materials
  by all-dry viscoelastic stamping}}.
\newblock \emph{\bibinfo{journal}{2D Materials}} \textbf{\bibinfo{volume}{1}},
  \bibinfo{pages}{011002} (\bibinfo{year}{2014}).

\bibitem{Renner1991a}
\bibinfo{author}{Renner, C.}, \bibinfo{author}{Kent, A.~D.},
  \bibinfo{author}{Niedermann, P.}, \bibinfo{author}{Fischer} \&
  \bibinfo{author}{L{\'{e}}vy, F.}
\newblock \bibinfo{title}{{Scanning tunneling spectroscopy of a vortex core
  from the clean to the dirty limit}}.
\newblock \emph{\bibinfo{journal}{Phys. Rev. Lett.}}
  \textbf{\bibinfo{volume}{67}}, \bibinfo{pages}{1650--1652}
  (\bibinfo{year}{1991}).

\bibitem{Sharvin1965}
\bibinfo{author}{Sharvin, Y.~V.}
\newblock \bibinfo{title}{{A possible method for studying Fermi surfaces}}.
\newblock \emph{\bibinfo{journal}{JETP}} \textbf{\bibinfo{volume}{48}},
  \bibinfo{pages}{984--985} (\bibinfo{year}{1965}).

\bibitem{Griffiths2005}
\bibinfo{author}{Griffiths, D.~J.}
\newblock \emph{\bibinfo{title}{{Introduction to quantum mechanics}}}
  (\bibinfo{publisher}{Pearson Education India}, \bibinfo{year}{2005}).

\bibitem{Brinkman1970TunnelingBarriers}
\bibinfo{author}{Brinkman, W.~F.}, \bibinfo{author}{Dynes, R.~C.} \&
  \bibinfo{author}{Rowell, J.~M.}
\newblock \bibinfo{title}{{Tunneling Conductance of Asymmetrical Barriers}}.
\newblock \emph{\bibinfo{journal}{Journal of Applied Physics}}
  \textbf{\bibinfo{volume}{41}}, \bibinfo{pages}{1915} (\bibinfo{year}{1970}).

\end{thebibliography}

\pagebreak
\widetext
\begin{center}
\textbf{\large Supplemental Materials: Spectroscopy of bulk and few-layer superconducting NbSe$_2$ with van der Waals tunnel junctions}
\end{center}
\setcounter{equation}{0}
\setcounter{figure}{0}
\setcounter{table}{0}
\setcounter{page}{1}
\makeatletter
\renewcommand{\theequation}{S\arabic{equation}}
\renewcommand\thesection{S\arabic{section}}
\renewcommand\thefigure{\textbf{S\arabic{figure}}}   
\renewcommand{\figurename}{\textbf{Supplementary Figure}}
\setcounter{secnumdepth}{1}

\section{Details of the measurement setup and the fitting process}

\begin{figure}[h]
	\centering
	\includegraphics[width=0.5\textwidth]{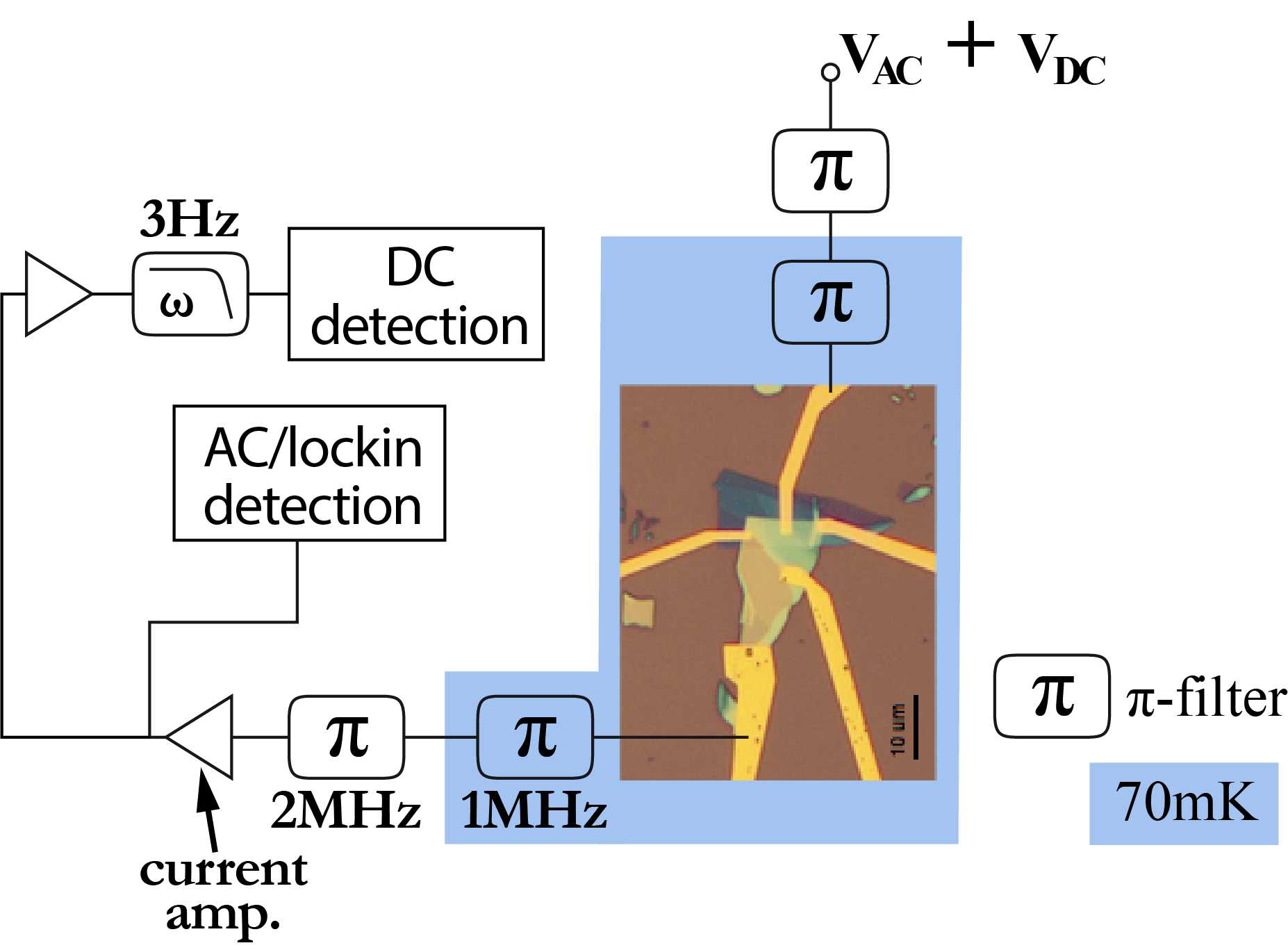}
	\caption[caption]{Detailed diagram of the measurement circuit used in the experiment.}
	\label{circuit}
\end{figure}

Figure~\ref{circuit} shows our measurement circuit in greater detail than was presented in the main text. All $\pi$-filters at low temperature have cutoff frequencies of 1MHz while those at room temperature have cutoff frequencies of 2MHz. The amplitude of the AC excitation $V_{AC}$ is 15$\mu$V in all the figures of the main text. Measurements at lower $V_{AC}$ showed that, between 2$\mu$V and 15$\mu$V, there was no discernible distortion of $G(V)$; the higher excitation voltage was thus chosen in order to have a better signal-to-noise ratio.

To fit the resulting dI/dVs to the SSM model as discussed in the main text, the data was horizontally shifted, to account for zero bias drift, and divided by a dI/dV curve taken at T > T$_C$. It was then symmetrized around zero bias, and fitted using least-squares method to the SSM model. The data shown in the figures, in the main text and in the supplementary, is either the original data as measured, or the data divided by a constant normalization factor, shifted horizontally by a constant bias.

\section{Thickness and structure of the tunnel barrier}

\begin{figure}[h]
	\centering
	\includegraphics[width=0.5\textwidth]{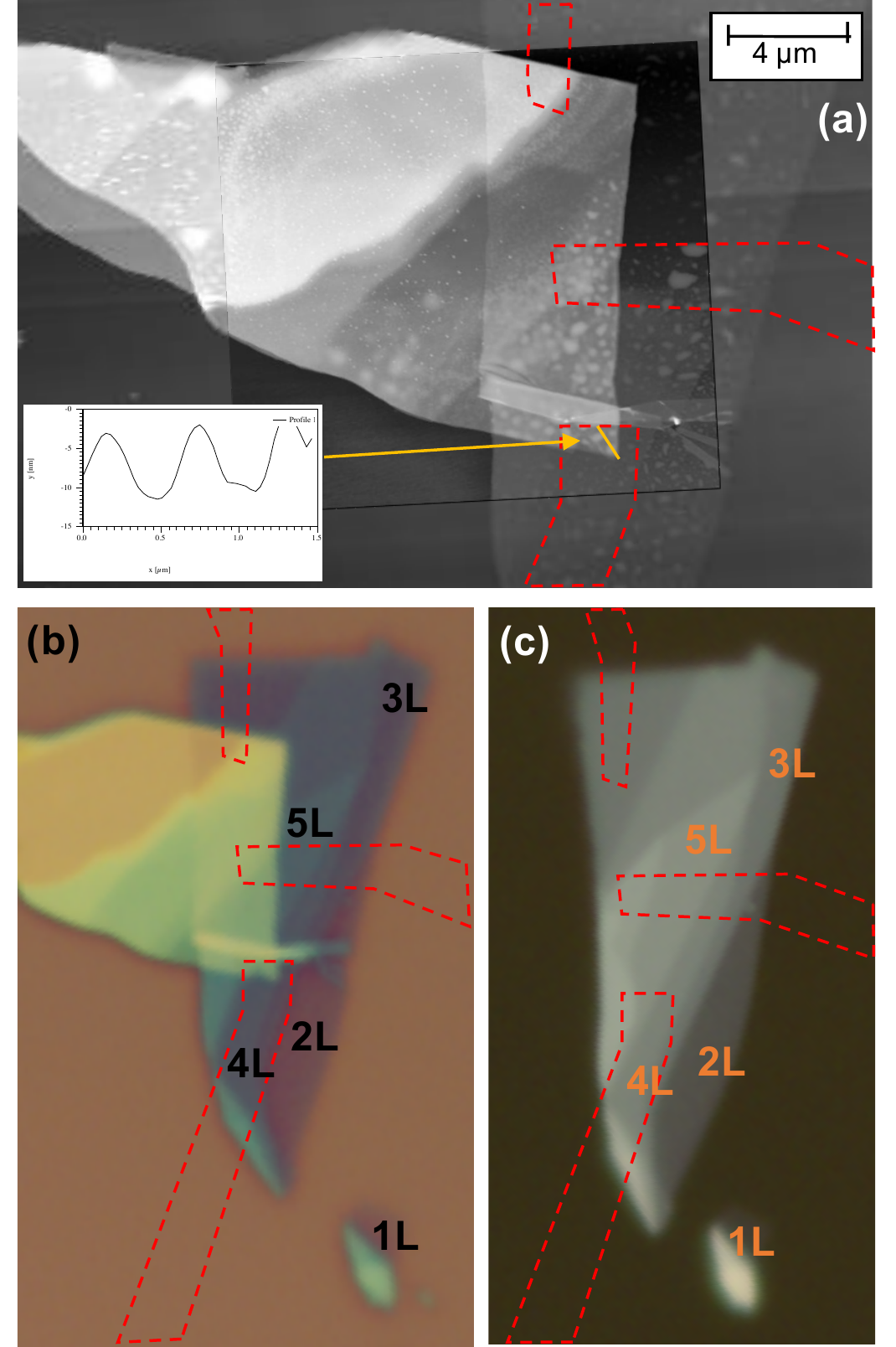}
	\caption[caption]{\textbf{AFM and optical images of the device. a,} AFM imaging of the bulk device discussed in the main text. Position of the tunnel electrodes marked in dashed-red. Inset: cross section of the solid blue line, showing the typical size of the dirt on the device. \textbf{b,} optical image of the two flakes prior to the deposition of the electrodes. Black numbers mark the number of layers observed, from 1 to 5. \textbf{c,} optical image of the \chem{MoS_2} flake on the PDMS prior to the transfer process. Orange numbers mark the number of layers observed, from 1 to 5.  }
	\label{layers}
\end{figure}
The high optical contrast between layers of different thickness of transition metal dichalcogenides (TMDs) allows easy identification of the thickness of the tunnel barrier. Figure \ref{layers} shows the optical image of the barrier on the PDMS immediately after it was exfoliated (panel c) and on top of the \chem{NbSe_2} flake after the transfer procedure (panel b). Both show clearly that the source electrode was deposited above a region consisting of 4 and 5 layer thick \chem{MoS_2}. As a result of exponential dependence of the tunnel current on the barrier thickness, only the 4 layer part of the junction is significant to the measurement. Hence we expect the effective junction area to be 1.6 $\mu m^2$ and the barrier thickness to be between 2.4 nm and 2.6 nm.

Contrary to the optical images, AFM does not provide a reliable measure of height between two different materials and cannot measure the thickness of the barrier. However AFM reveals some structures which are probably due to PDMS residue from the transfer process (panel a). A cross section of some of these features in the area of the junction shows height variation on the scale of 7 nm. The usual cleaning techniques of heat annealing cannot be used here due to the sensitivity of \chem{NbSe_2} to heat. The effect of this structure is most likely to reduce the effective area of the junction to the non-contaminated region. As discussed below, the effective area of the junction is of the same order of magnitude as the observed area, showing the robustness of this method to imperfections.

\section{Fitting of the SSM model to the spectrum of ultra-thin N\lowercase{b}S\lowercase{e}$_2$}
\begin{figure}
\includegraphics[width = 0.5\textwidth]{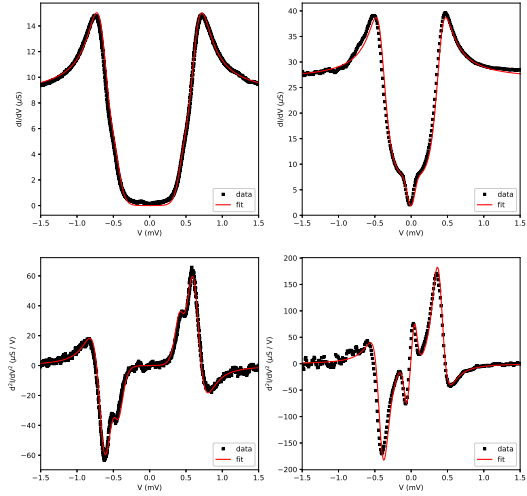}
\caption{\textbf{SSM model fits to 3- and 4-layer NbSe$_2$}, left: Differential conductance curve (top) and the second derivative (bottom) taken with the 4L sample and fits to SSM model. The fit to the lowest temperature curve is as discussed above. Right: Differential conductance curve (top) and the second derivative (bottom) taken with the 3L sample and fits to SSM model.  Fit parameters are given in table \ref{SSMtable} }
\label{ThinFits}
\end{figure}

The two band SSM can be used to fit spectrum obtained from 3- and 4- layers \chem{NbSe_2}. The fits are shown in figure \ref{ThinFits} and the fit parameters are given in table \ref{SSMtable}.  It is clear that the value of $\Delta_1^0$ decreases with a decreasing number of layers. In addition, the coupling constant $\Gamma_1$, which is associated with the rate of scattering of electrons from the band with the larger coupling, also decreases with decreasing number of layers. 

\begin{table}[h]
\centering
\begin{tabular}{ || c | c | c | c | c | c | c | c ||}
\hline  
  Thickness & $\Delta_1^0$ & $\Delta_2^0$ & $\Gamma_1$ & $\Gamma_2$ & $N_1$ & $N_2$ & T  \\
  \hline 
  Bulk & 1.24 & 0.29 & 0.41 & 1.32 & 1 & 0.11 & 0.44 \\
  4 layers & 0.65 & 0.42 & 0.14 & 0.03 & 1 & 0.03 & 0.5 \\
  3 layers & 0.4 & 0 & 0.09 & 0.09 & 1 & 0.04 & 0.32 \\
  \hline 
\end{tabular}
\caption{Summary of the fitting parameters to the SSM model}
\label{SSMtable}
\end{table}

The KZ model extends  the SSM model to describe the spectrum under the application of in-plane magnetic field. This requires the introduction of a depairing parameter $\Gamma^{AG}_i$ for each band. We show the value of $\Gamma^{AG}_i$ extracted from the model fit for each device in the table below. We stress that the observed change in the spectrum with magnetic field is very small in the case of the thin samples, and in the band with the larger gap of the bulk samples. Thus, the obtained values of $\Gamma^{AG}$ should be treated as bounds, rather then accurate numbers. Further study, conducted at higher fields would allow for a more precise estimation.

The depairing represented by $\Gamma^{AG}$ can originate from orbital depairing or from the interaction between the electrons' spin the applied magnetic field. Interpreting the depairing in terms of diffusive orbital effect allows us to extract the diffusion constant, $D$, as explained in the main text. Assuming a Fermi velocity $v_F = 10^5$ m/sec, typical to \chem{NbSe_2} $K$ bands, we can extract the elastic mean free path, $l_\textrm{mfp}$ and the time between scatterings, $\tau$. The mean free path was previously found to be in the range $l_\textrm{mfp}\approx$ 30-80 nm \cite{Renner1991a,Haven1980}. Table \ref{GammaAGtable} show that this interpretation gives plausible values for all bands except the second band in the 3L device. The values for the second bands in the bulk and 4L device are slightly higher than expected. This can be resolved by assuming a higher Fermi velocity, consistent with their identification as originating from the Se derived $\Gamma$ band.  In the 3L device, the depairing is too high to originate from orbital depairing, and has to be associated with spin-field interaction.

To interpret the depairing in terms of the interaction between the field and the spins, we note that away from the extreme paramagnetic limit, we should observe Zeeman splitting of the quasiparticle peaks. The effect can be suppressed by the spin-orbit interaction in one of the ways - spin flip during scattering process, that averages the projection of the spin on the magnetic field; and renormalization  of the in-plane magentic field by the effective out-of-plane field generated by the Ising spin-orbit coupling. The former effect is quantified by a typical time for spin flip, $\tau_{SO}$, whereas the latter is quantified by a spin orbit energy. Both enter the KZ model through  $\Gamma^{AG}$. Table \ref{GammaAGtable} shows that for the bands 1,2 of the 4 layer device and band 1 of the 3 layer device, $\tau_{SO}$ is shorter than the typical elastic scattering time, making this interpretation implausible.

\begin{table}[h]
\centering
\begin{tabular}{ || c | c | c | c | c | c | c | c ||}
\hline  
  Thickness & $\Gamma^{AG}$ [$\mu$ev/T$^2$] & $D$ [cm$^2$/sec] & $l_{\textrm{mfp}}$ [nm] & $\tau$ [fsec] & $\tau_{SO}$ [psec] & $\Delta_{SO}$ [meV]  \\
  \hline 
  Bulk, band 1      & 50     & 3      & 11   & 0.1   & - & -  \\
  Bulk, band 2       & 640  & 40  & 135     & 1.5 & - & - \\
  4 layers, band 1 & 0.8   & 6     & 18    & 0.2   & 0.08 & 8.2  \\
  4 layers, band 2 & 2       & 14   & 46      & 0.5  & 0.2 & 3.3  \\
  3 layers, band 1 & 0.5   & 6   & 20   & 0.2    & 0.05& 13  \\
  3 layers, band 2 & 6    & 75 & 250   & 2.7     & 0.6 & 1.1 \\
  \hline 
\end{tabular}
\caption{Summary of the fitting parameters to the KZ model and their possible interpretations.}
\label{GammaAGtable}
\end{table}

\section{Temperature dependent differential conductance}

\begin{figure}
\includegraphics[width = 0.5\textwidth]{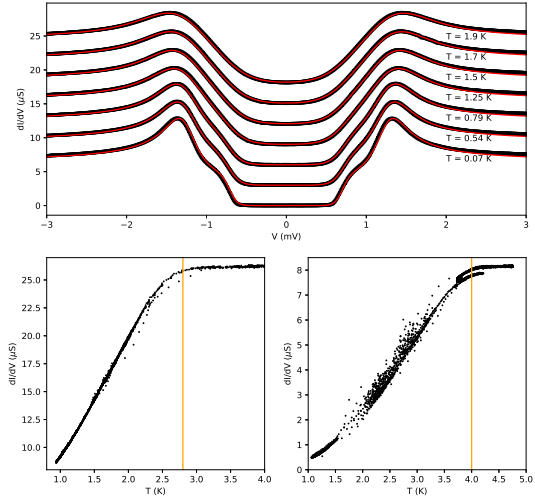}
\caption{\bf{Temperature dependent differential conductance, a.} Differential conductance curves taken with the bulk sample and fits to SSM model. The fit to the lowest temperature curve is as discussed above. To fit the higher temperature curves, only $T$ and $\Delta$ were allowed to change. \bf{b.} Zero bias conductance as a function of temperature for the bilayer sample. Orange line marks $T_c$. \bf{c.} Zero bias conductance as a function of temperature for the four layer sample. Orange line marks $T_c$.  }
\label{Tdependence}
\end{figure}

Figure \ref{Tdependence}a shows the differential conductance curves taken with the bulk sample at different temperatures. To fit the data, first the lowest temperature curve was fitted to the SSM model as discussed above. Then, the other curves were fitted to the model with the same parameters, where only the temperature and order parameters ($T,\Delta_1,\Delta_2$) were allowed to change. Best fit was obtained when the order parameters remained unchanged. Panels b and c show the zero bias conductance as a function of temperature for the three and four layer samples, respectively. This data was used to determine $T_c$, as transport data was unavailable. The critical temperature was defined by 5\% reduction in the conductance.

\section{Estimate of the barrier transparency}

We can estimate the transparency of our tunnel barrier $\mathcal{T}$ from the well-known expression from Sharvin~\cite{Sharvin1965}:

\begin{equation} \label{sharvin-eq}
G_N = \frac{2e^2}{h}\frac{k_F^2 A}{4 \pi}\mathcal{T},
\end{equation}

where $G_N$ is the junction conductance in the normal state, $A$ is the area of the junction, $k_F$ the Fermi momentum and $\mathcal{T}$ the average transmission of each conductance channel. We measure $G_N=$ 7$\mu$S for $A =$ 1.6 $\mu m^2$. $k_F$ in metals is usually $\sim 10^{10}~m^{-1}$ and it is about half this value in \chem{NbSe_2}. Taking the lower value, we get $\mathcal{T}\sim 3\times10^{-8}$.

We can make an independent estimate of $\mathcal{T}$ using the textbook WKB formula for a square barrier of thickness $d$ and height $U$ \cite{Griffiths2005}:

\begin{equation} \label{griffiths}
\mathcal{T} =\exp(-2d\sqrt{2m^*U}/\hbar)
\end{equation}

where $m^*$ is the effective mass of the electron in the barrier, here \chem{MoS_2}.

The gap of few layer \chem{MoS_2} at the $\Gamma$ point in the Brilloiun zone is on the order of 2eV, whereas the effective mass is generally a fraction of 1. Taking $U=$ 1eV, $m^*=m/2$ ($m$ being the bare electron mass), and $d$ in the range 2.4--2.6nm we find $\mathcal{T}\sim3\times 10^{-8}$--$6.5\times 10^{-9}$, consistent with the Sharvin estimate. 

We can make a more rigorous estimate of $U$ (and thus $\mathcal{T}$) by using Brinkman et al.'s result~\cite{Brinkman1970TunnelingBarriers} for the conductance across a trapezoidal barrier with diffuse boundaries, together with measurements of the high bias conductance of our junction:

\begin{equation} \label{brinkman-eq}
\frac{G(V)}{G(0)} = 1-\frac{A_0\Delta\phi}{16\bar{\phi}^{3/2}}eV+\frac{9}{128}\frac{A_0^2}{\bar{\phi}}(eV)^2 
\end{equation}\\

where $V$ is the voltage across the barrier, $\bar{\phi}$ is the mean barrier height, $\Delta\phi$ the barrier height difference on the two sides of the trapezoid, $\mathtt{d}$ the barrier width and $A_0 = 4\sqrt{2m^*}\mathtt{d}/3\hbar$. In these expressions, $\mathtt{d}$ is in units of \AA, while $\bar{\phi}$, $\phi$ and $V$ are in units of volts.

\begin{figure}[h]
	\centering
	\includegraphics[width=0.5\textwidth]{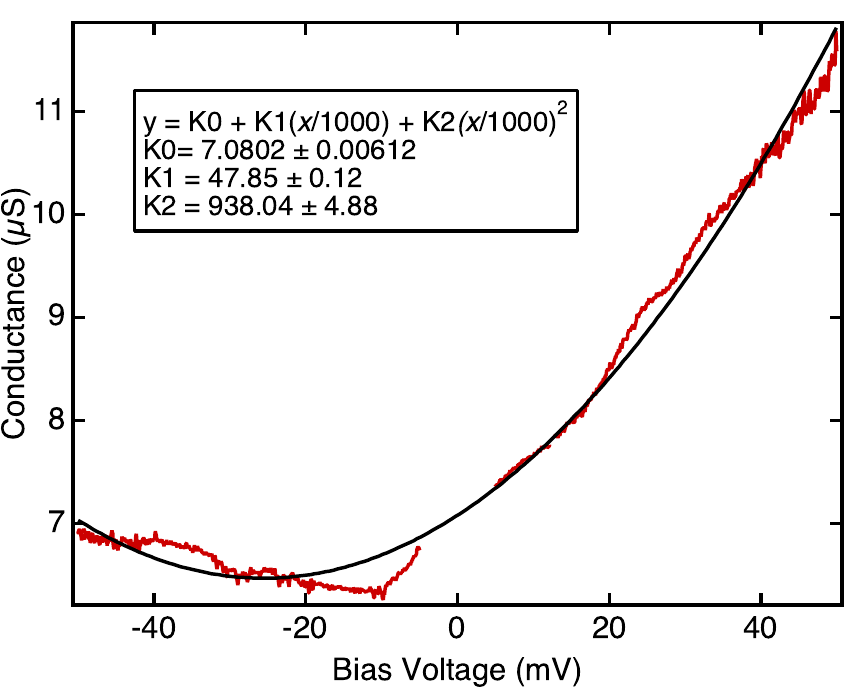}
	\caption[caption]{Conductance as a function of voltage at high biases (red) with a parabolic fit (black). The fit allows us to estimate our barrier height.}
	\label{background}
\end{figure}

Far from the Fermi level, the conductance of our junction indeed rises (Figure~\ref{background}). This rise is not perfectly parabolic and is likely due, in part, to factors other than barrier transparency and asymmetry. Therefore, fitting a parabola to the background, i.e. assuming that the rise is due almost entirely to the barrier, will give us a worst case scenario or minimum possible barrier height.

From the fit to our data to Equation~\ref{brinkman-eq} using $\mathtt{d}$ = 20\AA, we find $\bar{\phi}\sim$0.8V, not so different from what we assumed previously. If we use this, and $d = $ 2.4--2.6nm as before, $\mathcal{T}\sim2\times10^{-7}$--$5\times 10^{-8}$.

Considering all of the above, $\mathcal{T}$ is likely in the $10^{-8}$ range or close to it.

\end{document}